# Colloidal stability of tannins: astringency, wine tasting and beyond

D. Zanchi (1), C. Poulain (1,2), P. Konarev (3), C. Tribet (4), D.I. Svergun (3)

drazen@lpthe.jussieu.fr

(1) Laboratoire de Physique Théorique et Hautes Energies, 4 Place Jussieu, BP 126 - 75252 Paris Cédex 05 - France.

(2) Laboratoire Materiaux et Phénomènes quantiques, Université Paris Diderot, 10 rue Alice Domon et Léonie Duquet, 75013 Paris – France.

(3) European Molecular Biology Laboratory, Hamburg Outstation, Notkestrasse 85, 22603 Hamburg, Germany.

(4) Physico-Chimie des Polymères et Milieux Dispersés, CNRS UMR 7615, ESPCI, 10, rue Vauquelin, 75231 Paris Cédex 05, France.

**Abstract**

Tannin-tannin and tannin-protein interactions in water-ethanol solvent mixtures are studied in the context of red wine tasting. While tannin self-aggregation is relevant for visual aspect of wine tasting (limpidity and related colloidal phenomena), tannin affinities for salivary proline-rich proteins is fundamental for a wide spectrum of organoleptic properties related to astringency.
Tannin-tannin interactions are analyzed in water-ethanol wine-like solvents and the precipitation map is constructed for a typical grape tannin. The interaction between tannins and human salivary proline-rich proteins (PRP) are investigated in the framework of the shell model for micellization, known for describing tannin-induced aggregation of β-casein. Tannin-assisted micellization and compaction of proteins observed by SAXS are described quantitatively and discussed in the case of astgringency.

**Introduction**

Tannins are phenolic compounds that can interact with proteins and in some cases precipitate them out [1,2]. They are large molecules whose molecular weight can vary between 300 Da and more than 50 kDa. They contain aromatic rings and OH groups that are available for binding to proline sites of proteins by H-bond and hydrophobic interactions. Structure of vegetal condensed tannins is shown in figure 1. From the biological point of view, tannins are part of the plant's defence system against bacteria, viruses and higher herbivores. For instance, when attacking a fruit, bacteria secrets pectinazes and cellulazes in order to break the cell walls and to accede the juice. Tannins liberated in this process inhibit the enzymes by inducing their aggregation. Similarly, when higher herbivores ingest an unripe fruit, tannins inhibit the digestive enzymes reducing digestibility of the food. To tackle this problem herbivores developed proline-rich proteins (PRP) in saliva, which bind to tannins and reduce their activity before ingestion. Tannin-induced aggregation of salivary proteins produces physical changes of salivary mixture, in particular a radical drop of the viscosity. This is believed to be the basic mechanism of astringency which is therefore not a taste but merely a tactile sensation of dryness in the mouth [3].

More generally, the control of protein aggregation is an issue of broad and interdisciplinary interest. Protein aggregates have variety of applications in technology and in biomaterials design [4]. On the other hand, protein aggregation and precipitation are well known for causing many undesirable consequences including a wide range of highly debilitating and prevailing diseases [5]. Tannins, by sticking together and/or by bridging proline sites of proteins are efficient protein aggregators whose practical use can be envisaged. Namely, tannin-protein molecular assemblies either remain soluble, or just precipitate, or, in some specific microscopically controlled conditions, form interesting materials as tanned silk, mussel byssus, insect cuticle, etc [2].

In this paper we point out the importance of tannins as efficient and site specific tool for inducing micellization and radical compaction of unfolded native proteins that contain prolines. We concentrate to one particular system, the red wine. Tannins contribute in the body, in the structure and in the robe of the red wine. Upon ageing tannins become velvety, silky, smooth, fine-grained, supple and well-integrated. If the wine is too young its tannins are green, harsh, rough, coarse, firm, chewy, tough and hard. When the wine is too old its tannins are dusty and the wine is said to have dried out. It is curious that red wine tannins upon ageing form aggregates which are typically metastable in the beginning but finally start to form hazes and precipitates which are often considered as defaults. Our present analysis addresses two questions: (i) is it possible to relate the tannins' "quality" with the wine elaboration protocol? and (ii) what is the molecular basis of astringent sensation and why it can take different organoleptic forms?

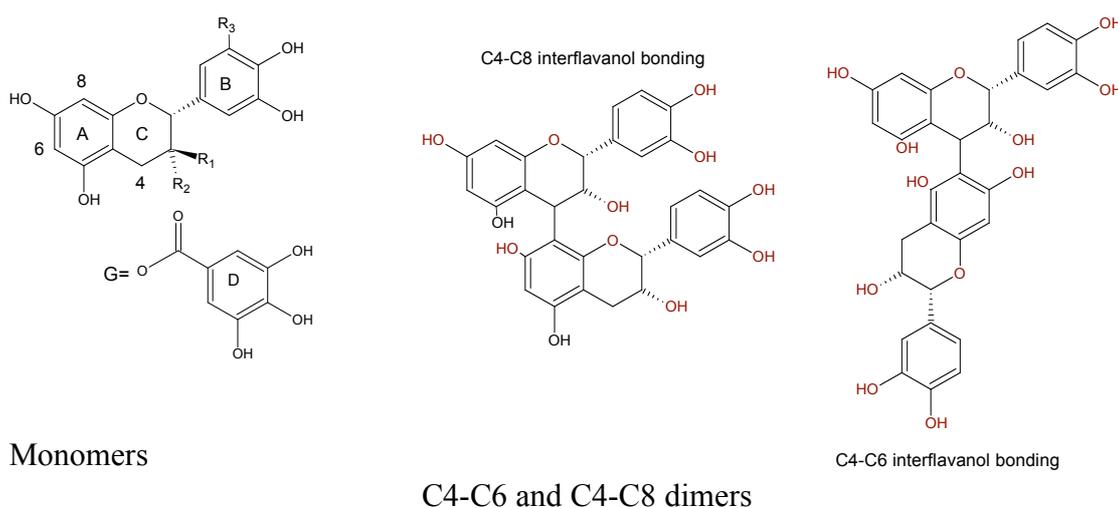

Monomers                    C4-C6 and C4-C8 dimers

Figure 1 Structures of condensed tannins. The flavane kernel is made of two aromatic rings (A) and (B) and a saturated pyrane ring (C). Grape condensed tannins are made of catechin (R1 = OH, R2 = R3 = H), epicatechin (R1 = R3 = H, R2 = OH), epigallocatechin (R1 = H, R2 = R3 = OH) and epicatechin gallate (R1 = R3 = H, R2 = G). Monomers are linked through C4-C6 and C4-C8 interflavanol bonds. The smallest tannin is epigallocatechin gallate or EGCG (R1 = H, R2 = G, R3 = OH).

We first study the physical state of tannins in water-ethanol model mixtures and then we analyze the SAXS results obtained on mixtures of human salivary prolin-rich protein (PRP) II-1 and tannins in model water-ethanol solution. Our result help in elucidating 1) self aggregation and precipitation of tannins in red wines and 2) the mechanisms of astringency at nanometer scales. We quantify the aggregation, the morphology and the internal structure of tannin-protein nano-assemblies. In particular we concentrate on the phenomenon of tannin-assisted compaction of protein micelles. Similar phenomena are observed when small tannins (smaller than about 2kDa) are brought in contact with β-casein [6], skim milk caseins [7] and gelatines [8]. Thus tannin-assisted micelization and compaction of unfolded native proteins seems to be general effect, and not just an interesting anecdote about a glass of red wine.

In second section we review results on tannins in water-ethanol solutions. Third section contains new results: analysis of SAXS data on tannin - salivary protein water solutions and theoretical model of astringency. Last section contains conclusion and perspectives.

**Tannins in water-ethanol solution**

Colourless, bitter and extremely astringent condensed tannins and deeply red-to-violet anthocyanins are extracted from the grape seeds and skins, mostly by alcohol, during wine elaboration. Upon further ageing condensed tannins and anthocyanins continuously re-arrange, polymerize and co-polymerize, branch and break. All products, often macromolecules, are called the red wine tannins. Their chemical evolution continues all the life-long of the wine, until the evening the bottle is opened. But this is only the chemical part of the story. Red wine polyphenols or tannins have important hydrophobic interactions together with strong hydrogen bounding tendencies through their hydroxyl groups. By means of hydrophobic interactions some of tannins are likely to associate making thus the red wine a meta-stable colloidal dispersion. Tannin aggregation and related colloidal phenomena take therefore an important part in wine haze formation, in filtration and stabilization techniques [9].

Tannin-tannin interactions in wine-like solution [10] are best appreciated on the precipitation map shown in figure 2. It was obtained for a grape seed condensed tannin with average polymerization number DP=11, which is not far from the value in real new red wines. The solvent conditions were adjusted by sodium tartarate to pH = 3.4 and to the ionic strength of 10mM.

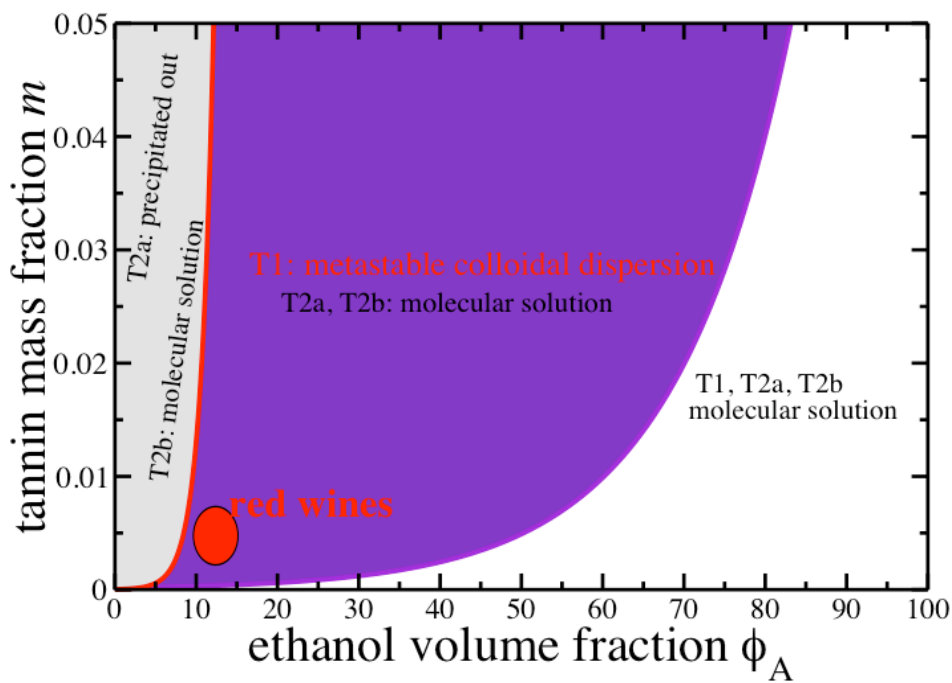

Figure 2 Nucleation and precipitation thresholds for grape seed tannin mDP11 in tartarate buffer - ethanol mixture; ionic strength at $\phi_A$=0.12 is 10mM and pH=3.4, which corresponds to a typical wine composition. The nature of the different fractions T1, T2a and T2b are explained in the text.

By decreasing ethanol content (adding water) there is a sequence of precipitation stages. The first stage has a threshold at about $\phi_A = 0.6$ for 10g/L of tannin, and goes on all the way to the typical wine ethanol content $\phi_A = 0.12$. During this stage, T1 population of tannins forms colloidal particles that remain dispersed in the solution. The second stage occurs at a lower ethanol content. During this stage, another, much larger population T2a forms colloidal particles that aggregate readily and separate out from the solution. The remaining tannin fraction T2b remains in solution regardless of ethanol content and ionic strength.

T1 population of tannin macromolecules can contain from 1 to more than 10 percent of total tannin mass. It is *hydrophobic*, since its nucleation occurs as soon as there are enough water molecules to create a network of hydrogen bonds that resembles that of liquid water. T2 is the majority population of tannin macromolecules that remains in solution all the way to the wine ethanol content, $\phi_A = 0.12$. This population is *hydrophilic*, since it remains highly soluble in solvents where the hydrogen bond network is nearly that of liquid water. T2a is the subpopulation that precipitates at still lower ethanol content and moderate ionic strength, and T2b is the subpopulation that always remains in solution.

It is surprising that a tannin fraction with a well specified polymerization degree contains in fact three well distinguished fractions. A key element for the comprehension of this property is found in recent experiments on tannins aged in controlled conditions [11]. Results indicate that hydrophobic tannin fraction T1 are in fact molecules that were affected by oxidation without changing their molecular weight by more than several units, corresponding to just additional inter-flavane intra-molecular condensations. Resulting molecules are in more extended conformations due to geometrical constraints and consequently their hydrophobic surfaces are more exposed to solvent.
On the other hand, the fractionation of hydrophilic T2 tannins into T2a and T2b is not yet elucidated. Two main effects can give us some indications in this respect: 1) strong sensibility of T2 fractionation to the buffer (for example tartarate buffer is very efficient, while simple inorganic salts are less) and 2) T2a solubility boundary can be understood only if we assume that preferential solvation (strong solvent heterogeneities in the solvation layer) takes place. These two observations indicate that T2 fractionation is determined by specific tannin - organic acid complexation and strongly heterogeneous interactions between tannin and water-ethanol solvent. We must also remember that a tannin fraction with single polymerization degree can contain a large number of different conformers and branched structures.

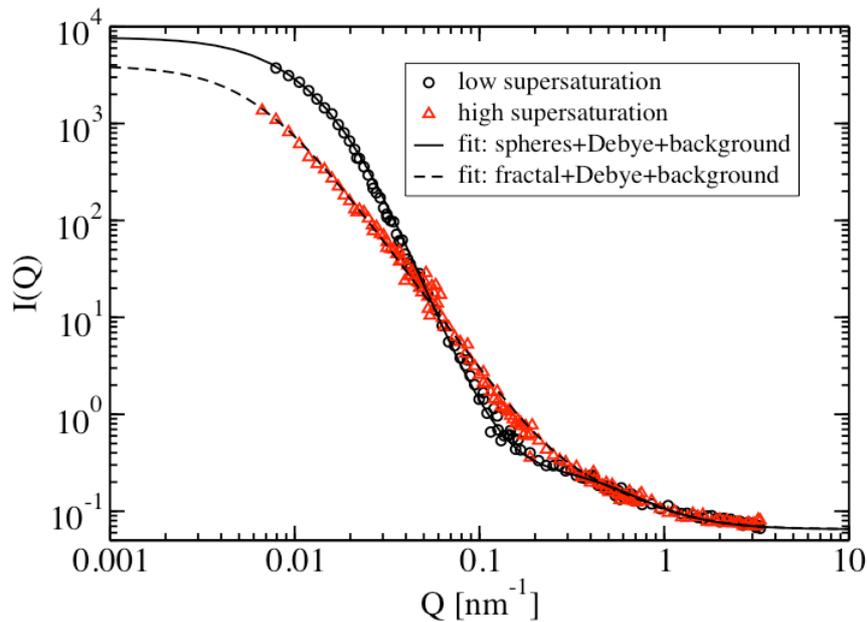

Figure 3 Comparison of dispersions created through different mixing procedures, with the same final composition: ethanol volume fraction $\phi_A = 0.12$ and tannin concentration 5 g/L. Circles: Dispersions created by addition of aqueous solution into the ethanolic tannin solution (low supersaturation); the fit is the calculated scattering curve of polydisperse spheres with an average radius R = 170 nm co-existing together with tannin polymers in solution (Debye formula fit with $R_g$=2.9 nm). Triangles: Dispersions created by direct mixing of the ethanolic tannin solution into a large volume of aqueous solution (high supersaturation); the fit is the calculated scattering curve of fractal aggregates with an average radius of gyration $R_g$ = 320 nm and a fractal dimension $d_f$ = 2.6, together with polymeric tannins in solution (Debye formula with $R_g$=2.1 nm). Incoherent background scattering intensity is 0.065 cm$^{-1}$.

From practical point of view it is important that tannin – tannin interactions alone may be sufficient to cause the formation of colloidal dispersions of tannins in water – ethanol solutions and that these colloids can be stabilized for T1 fractions. If the sample is prepared in a way that mimics the traditional wine making procedure, metastable colloidal aggregates are typically spherical objects with radius between 100 and 200 nm and with well defined Porod decay, as shown in figure 3. All samples are prepared by mixing stock solution of tannin in ethanol with water buffer. Slow adding of stock solution into buffer implies that the phase boundary is crossed in a continuous way, so that tannins start to nucleate in low supersaturation conditions. This procedure is similar to slow tannin extraction/evolution in wine making process. Resulting aggregates are large and have smooth surface because in low supersaturation conditions only large nuclei survive while small ones, together with all irregularities on the surfaces are immediately re-dissolved.

Alternative route to prepare samples is by direct and rapid mixing of stock solution and water buffer. That way the tannins are immediately in high supersaturation conditions, which implies rapid aggregation at all scales starting from smallest ones. Resulting aggregates are fractal, as it can also be seen on SANS profile in figure 3. This later case is not expected to occur in traditional wine making, but it can occur in cases when tannin is added into wine artificially.

Colloidal stabilization of T1 tannin is another intriguing phenomenon. Tannins are not charged at acid pH and one would naively expect that aggregation continues until the phase separation. The mechanism that prevents colloidal particles to aggregate is in fact the already mentioned complexation between tannins and carboxylic acids in the solvent. It has been shown in [10] that already very low concentration of organic acid (less than one acid per 100 tannins) is sufficient for this stabilization by simple electrostatic mechanism. Indeed, microelectrophoresis experiments on tannins in tartatare buffers, reported also in [10], show that substantial fraction of tannins acquire charges by complexation with tartarate ions.

**Tannin-assisted compaction of proteins and astringency**

Systematic research of astringency as a function of structure and molecular mass of tannin is a vast project. In the present study we analyze the effects of several rather short (M < 4.5 kDa) well defined water soluble (T2 type) tannin fractions on a single human salivary proline-rich protein (PRP) II-1.
It is a natively unfolded protein. It was obtained by recombinant pathway [12]. For its high proline content (51 of 159 aminoacids) and amphiphilic block character this protein is believed to be the most relevant for astringency. All proline sites are periodically placed, mostly in groups of 3 or 5, on the dominant hydrophilic side (sites 25-159) of the sequence, and are accompanied with six regularly disposed glycosylated segments.

Experiment [13] at rather high ionic strength (100mM NaCl) and pH=4 showed that aggregates are formed when tannins are added. Typical scattering profile on tannin-induced aggregates is shown in figure 4. Upon increasing tannin content at constant protein concentration, or when protein concentration is increased at constant tannin/protein ratio $t$, forward SAXS intensity increases while high-$Q$ part of the scattering profiles remains unchanged. These are indications that tannin induces aggregation of proteins. In order to understand quantitatively these results we used three different procedures for data analysis. First we assumed that proteins are bound into star-like structures by just one sticking point in the center. Using the form factor for star-like polymers [14] we fit data in satisfactory way as shown in figure 4. Second, we assume that proteins are bound by a large number of sticking points. The resulting structure can be seen as a hierarchical one. We therefore used the Beaucage unified function [15] and fit the data well again. Finally, we used *ab initio* procedures from the ATSAS package [16] for structure reconstruction.

Structures corresponding to two first models are shown in inset of figure 4. The Beaucage function assumes that polymers are cross-linked by stickers within an aggregate with larger gyration radius. Shorter gyration radius corresponds in that case to the distance between stickers (nods). Our point is that both fits, the one by Beaucage formula and the one by star polymer form factor, reveal the same overall gyration radius $R_g$ and average aggregation number N. Therefore these quantities are considered as invariants to the fitting procedure. The *ab initio* calculation confirmed these values and also provides a closer geometrical information about individual proteins and aggregates, as shown on right panel of figure 4. Single proteins are unfolded with a coil-like shape (Gaussian $Q^{-2}$ scaling) with a gyration radius of 12 nm. The tannin-induced protein aggregates keep the same gyration radius. For that reason we call them micelles. Within the micelle proteins are organized in a compact way. This compaction is the key for understanding of the molecular mechanisms of the astringency, which is a tactile sensation produced in mouth by reduction of lubricant property of saliva. In fact, the fall of viscosity is simply a consequence of the reduction of the effective volume fraction of protein. Dependences of the average aggregation number $N$ on tannin content $t$ and on protein concentration $P$ obtained by these fitting procedures are shown in figure 5.

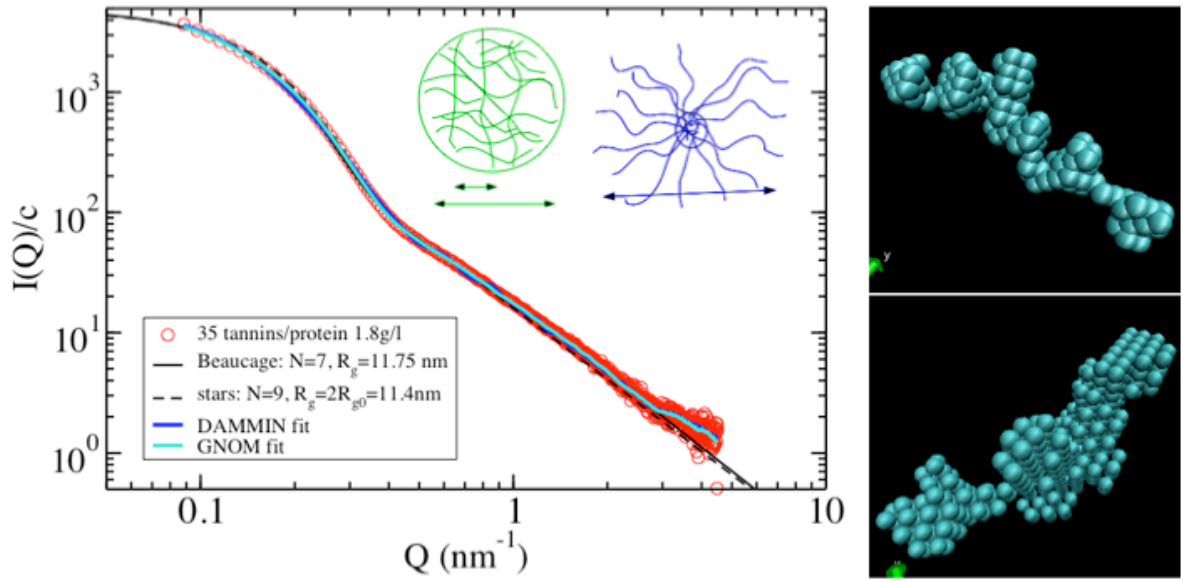

Figure 4 Left: Fits of the SAXS intensity for human salivary proline-rich protein – tannin mixture at t=35 tannins per protein with protein concentration of 1.8 g/L. Solid line: Beaucage unified function for hierarchical structures [15], dashed line: star polymer model [14]. The essential feature is that both fits are obtained assuming that the whole scattering profile is dominated by protein-tannin micelles. Dark bold solid line: fit using DAMMIN procedure of ATSAS package for *ab initio* structure reconstruction [16]. Clear bold solid line: fit using GNOM procedure of ATSAS. Right: corresponding reconstructed models (results of DAMMIN); upper model is a single protein (t<5) , lower model is the protein-tannin aggregate containing nine proteins corresponding to t=35 tannins/protein. Gyration radius of the two structures is the same, of about 12 nm.

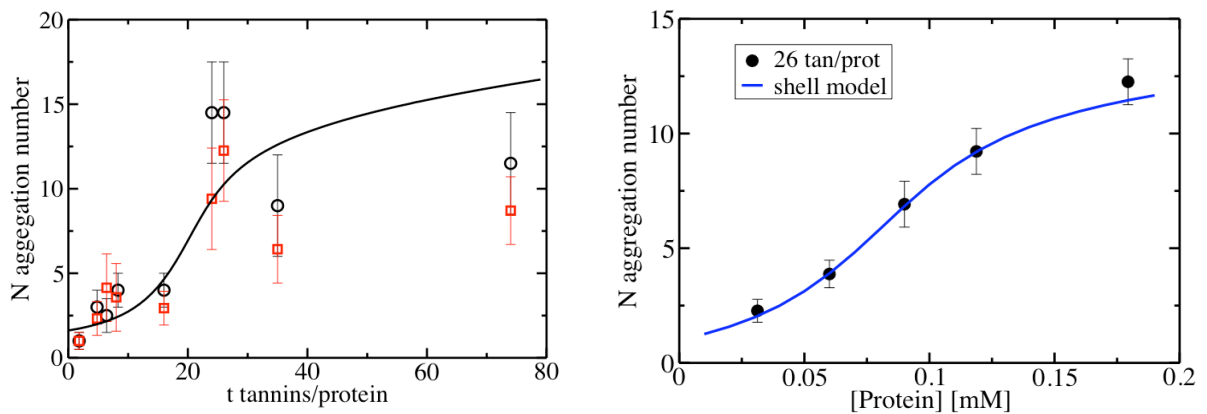

Figure 5 (a): Aggregation number of tannin-protein micelles N as function of tannin/protein ratio t at constant protein concentration of 1.8 g/L. (b): N as function of the protein concentration at constant t. Experimental points are obtained from fits of the SAXS data as explained in text. Solid lines are results of our theory based on the shell model.

From above analysis we found three remarkable facts: (i) linear dimensions of the aggregates are the same as the size of individual protein and do not depend on aggregation number, (ii) aggregation is induced by tannins above t=5 tannins per protein and (iii) the aggregation number saturates at higher tannin or protein content. These observations indicate that the relevant mechanism is a tannin-assisted micellization, as in the case of β-casein-tannin mixture [6]. We therefore use the same theory, which is based on the shell model [17] modified to include the effect of tannins as effective enhancers of hydrophobic interactions between proline sites [6].

The shell model is based on two plausible assumptions: (i) The aggregation number n of micelles cannot be greater than some maximal value $n_>$, (ii) the association rate of two free molecules is lower by a cooperativity parameter $f$ than the association rate of a free monomer and a molecule within an aggregate. This model calculates the aggregation number distribution for a cascade of accretion-dissociation processes given by

$$A_0 + A_{i-1} \leftrightarrow A_i \qquad /1/$$

where $A_0$ symbolizes a single free protein, while $A_i$ is a cluster composed of $i+1$ proteins.

In the presence of small tannins, the free energy difference G needed to bring together two solvated molecules is given by

$$G = G_0 + \gamma \alpha t \qquad /2/$$

where $G_0$ is the association free energy for pure protein, $\gamma$ is the free energy of protein-protein binding per bound tannin and $\alpha$ is the fraction of tannins bound to proteins.
The later, $\alpha$=[Prol.$_\nu$ Tan.]/[Tan.], is found from the equilibrium described by the dissociation constant of tannin-(proline)$_\nu$ complex $K_{pt}$=[Tan.][Prol. $_\nu$]/[Prol.$_\nu$ Tan.] 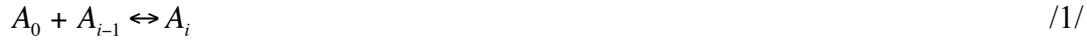 where each tannin binds to $\nu$ prolines. Results under consideration correspond to tannin epigallocatechin-gallate (EGCG) which binds just two prolines: $\nu$=2.

Both sets of data N(t) and N(P) (see figure 5) can be fitted by the shell-model assuming the upper cut-off $n_>$=30, cooperativity parameter $f$=0.05, bare protein dimerization free energy $G_0$=-18.5 kJ/mol, a rather high effective protein-tannin dissociation constant $K_{pt}$=0.1 mol/L, the fraction of tannin-available prolines of only 35 of 51 and protein-protein binding free energy per bound tannin $\gamma$=-4 kJ/mol.

It is instructive to compare these data with the known results for tannin - β-casein micelles [6]. The upper cut-off $n_>$=30 is low compared to its value of 175 found for β-casein. So far, we can only speculate about the reasons of this difference. Probably it is related to the fact that micelles do not have circular symmetry. Instead, proteins form brick-like stacks or strand-like structures, as the *ab initio* reconstruction suggests (figure 4). This implies that $n_>$ is related to the maximal stacking of the PRP-tannin complex and not to the closest packing of amphiphilic chains into a core-crown spherical micelle (as it is the case with β-casein). Indeed, the amphipilic block character of the PRP II-1 is less ample than the one of the β-casein, which makes core-crown structure more plausible for β-casein than for the PRP.
The parameter $f$=0.05 implies a moderate cooperativity: the probablity to form an isolated dimer is 5% of the probability to add a protein (to a specific site) into an aggregate. For comparison, the value $f$=0.15 for β-casein is known and corresponds to a rather large micellization "transition" and broader distribution of aggregation number.
A rather high bare protein dimerization free energy $G_0$=-18.5 kJ/mol implies that the CMC of the PRP alone is of the order of 1mM (more than 10g/l or volume fraction of about 1%). At concentrations of several grams per litter micelles are formed only in presence of tannins. On the other hand the β-casein associates spontaneously well below. Its CMC is of the order of 10 μM ($G_0$=-25.4 kJ/mol), and tannins increase the aggregation number of already existing micelles.
Another interesting information that our theory provides is the fraction of proline sites available for interaction with tannins. This number is only 35 of 51 prolines because in PRP chain the prolines appear in groups. When one of prolines from the group binds a tannin, the neghboring sites are less available because of geometrical constraints. In the case of β-casein all prolines are available because most of them are single. Finally, the protein-protein binding free energy per bound tannin $\gamma$=-4kJ/mol is close to the value of -5kJ/mol found for β-casein. It corresponds to a typical hydrophobic energy and justifies our interpretation that *tannins act as effective enhancers of hydrophobic interactions between proline sites.*

# Conclusion

In the present work we point out the relevance of tannins and their interaction with proteins for wine tasting and beyond, for biophysics and even for biomaterials design.

Within a tannin fraction with specified average molecular mass, three populations of macromolecules have been identified: T1, which forms colloidal metastable particles when the ethanol content of the solvent is brought below about 60%, T2a, which phase separates below about 10% of ethanol and at moderate ionic strength, and T2b, which remains soluble regardless of ethanol content. Upon oxidative ageing tannin polymers become progressively more hydrophobic and consequently less soluble. In that way the fraction T1 increases, while the fractions T2a and T2b get reduced. This means that T1 are "aged" tannins while T2 are "fresh" tannins. The internal structure of the colloidal particles of tannins T1 depends on the sample preparation procedure. Samples prepared in weak supersaturation conditions (similar to slow tannin extraction/evolution in real wines) form compact aggregates with sharp surface with 100-200 nm in radius. On the contrary, if samples are prepared in a way that tannins are injected in strong supersaturation conditions, fractal aggregates are formed.

In order to understand the physical mechanisms of astringency, which is closely related to tannin-induced aggregation of proteins in saliva, we explored what happens when T2 tannins come in contact with natively unfolded proline rich human salivary proteins. We find that tannin reduces the effective volume ratio of protein by sticking together proline sites. In that way the viscosity of the system is reduced. This reduces lubrication in mouth and produces an astringent sensation. The strength of this effect together with the size and the structure of tannin-protein aggregates determine the texture of the salivary mixture relevant for the wine tasting.

We believe that our findings go well beyond a comprehension of just wine tasting. Namely, tannin assisted micellization of unfolded proteins containing proline is probably a general mechanism by which living organisms protect their folded proteins from being inhibited by tannins from food. Interestingly, in most of the known cases the density of complex tannin-protein micelles approaches the folded protein density.

Relevance of the tannin length in the present context is still unclear. Namely, it has been shown that long tannins (DP=28) precipitate β-casein micelles, while short ones increase the stability of micelles. The hydrophilic corona of β-casein micelles is the essential element in this respect. In the case of salivary proline-rich proteins no equivalent effects were found: all tannins from monomer to 15-mer form micelles with the protein, with well defined maximal aggregation number. The core-crown structure of these micelles is not obvious: they look more like irregular compact blocks with flat fragments. Further research is envisaged in order to explore how the tannin length and structure determine its activity towards specific proteins.

From the practical point of view, mixed micelles between intrinsically unfolded proteins (β-casein or human PRP) and tannins can be used as tannin transporters (or *vectors*): both proteins act in a way to bind together, within a micelle, a large number of tannins hiding them from the aqueous solvent.
Finally, the use of tannins in biomaterial design is another perspective. Our findings show that tannins can indeed be used as efficient site-specific tool for controlled protein aggregation in bulk liquid environment. We are convinced that in more controlled conditions at micro- and nanometer scale new and spectacular tannin-assisted protein association modes are yet to be discovered.

**Acknowledgments:** We thank Theyencheri Narayanan for reading the manuscript. We acknowledge the support of the European Community - Research Infrastructure Action under the FP6 "Structuring the European Research Area Programme", contract number: RII3-CT-2004-506008. Laboratoire de Physique Théorique et Hautes Energies is Unité Mixte de Recherche du CNRS UMR 7589, Universités Denis Diderot – Paris 7 and Pierre et Marie Curie - Paris 6.

**References**

[1] Hemingway R W and Laks P M, (eds.) 1992, *Plant Polyphenols* (Plenum Press: New York)

[2] E. Haslam, *Practical Polyphenolics: from structure to molecular recognition and physiological action*; Cambridge University Press, 1998.

[3] Jobstl E, O'Connell J, Fairclough J P A and Williamson M P 2004, Biomacromolecules vol 5 p 942-949

[4] Zhang S 2003, Nat. Biotehnol. vol 21 p 1171

[5] Dobson C M 2003, Nature vol. 426 p 884

[6] Zanchi D, Narayanan T, Hagenmuller D, Baron A, Guyot S, Cabane B, Bouhallab S 2008, Europhys. Lett. vol 82 p 58001

[7] A. Shukla, T. Narayanan and D. Zanchi, *Interaction between casein micelles and tannins probed by synchrotron SAXS*, to be published

[8] D. Zanchi, A. Baron and S. Guyot, *Tannin-induced folding of gelatins*, in preparation

[9] Waters E J, Peng Z, Pocock K F, Jones G P, Clarke P and Williams P J 1994, J. Agric. Food Chem. vol 42 p 1761-1766

[10] Zanchi D, Vernhet A, Poncet-Legrand C, Cartalade D, Tribet C, Schweins R and Cabane B 2007, Langmuir vol 23 p 9949

[11] Zanchi D, Guyot S, Konarev P, Tribet C, Baron A and Svergun D I, *Rigidity, conformation and solvation of fresh and aged tannin macromolecules in water-ethanol solution,* to be published

[12] Pascal C, Bigey F, Ratomahenina R, Boze H, Moulin G and Sarni-Manchado P 2006, Protein Expression and Purification vol 47 p 524-532

[13] Cabane B, Zanchi D, Pascal C, Vernhet A, Poncet-Legrand C, Konarev P and Svergun D I 2006, EMBL report *Interactions of proteins with tannins*, unpublished

[14] Pedersen J S 2002, in *Neutrons, X-rays and Light Scattering methods Applied to Soft Condensed Matter*, eds P Lindner and Th Zemb (North Holland)

[15] Beaucage G 2004, Phys. Rev. E vol 70 p 031401

[16] Konarev P V, Petoukhov M V, Volkov V V and Svergun D I 2006, J. Appl. Cryst. vol 39 p 277–286

[17] Mikheeva L M, Grinberg N V, Grinberg V Ya, Khokhlov A R and de Kruif C G 2003, Langmuir vol 19 p 2913